\documentclass[%
 aip,
 amsmath,amssymb,
preprint,%
]{revtex4-1}

\DeclareSymbolFont{matha}{OML}{txmi}{m}{it}% txfonts
\DeclareMathSymbol{\varv}{\mathord}{matha}{118}
\usepackage[normalem]{ulem}

\usepackage{graphicx}% Include figure files
\usepackage{dcolumn}% Align table columns on decimal point
\usepackage{bm}% bold math
\usepackage[utf8]{inputenc}
\usepackage[T1]{fontenc}
\usepackage{mathptmx}
\usepackage{etoolbox}

\usepackage[T1]{fontenc}
\usepackage[utf8]{inputenc}
\usepackage[absolute,overlay]{textpos}
\usepackage{hyperref}
\usepackage{graphicx}
\usepackage{dcolumn}
\usepackage{xcolor}
\usepackage{siunitx}
\usepackage[all]{nowidow}
\usepackage{commath} % \abs{} -> |a|; \norm{] -> ||a||
\usepackage{braket}  % \braket{} -> <a>
\usepackage{lipsum} 
\usepackage{import}
\usepackage{color}
\newcommand{\blu}[1]{{\color{black}{#1}}}

\newcommand*{\citen}{}% generate error, if `\citen` is already in use
\DeclareRobustCommand*{\citen}[1]{%
  \begingroup
    \romannumeral-`\x % remove space at the beginning of \setcitestyle
    \setcitestyle{numbers}%
    \cite{#1}%
  \endgroup
}

\makeatletter
\def\@email#1#2{%
 \endgroup
 \patchcmd{\titleblock@produce}
  {\frontmatter@RRAPformat}
  {\frontmatter@RRAPformat{\produce@RRAP{*#1\href{mailto:#2}{#2}}}\frontmatter@RRAPformat}
  {}{}
}%
\makeatother

\begin{document}

\preprint{AIP/123-QED}

\title[]{Liquid Thread Breakup and the Formation of Satellite Droplets}
% Force line breaks with \\
\author{Lu\'is H. Carnevale}
\affiliation{Institute of Physics, Polish Academy of Sciences, Al. Lotnik\'ow 32/46, 02-668 Warsaw, Poland}
\author{Piotr Deuar}%
% \email{deuar@ifpan.edu.pl}
\affiliation{Institute of Physics, Polish Academy of Sciences, Al. Lotnik\'ow 32/46, 02-668 Warsaw, Poland}
\author{Zhizhao Che}
% \homepage{http://www.Second.institution.edu/~Charlie.Author.}%
\affiliation{%
State Key Laboratory of Engines, Tianjin University, 300350 Tianjin, China%\\This line break forced% with \\
}%
\author{Panagiotis E. Theodorakis*}%
 \email{panos@ifpan.edu.pl}
\affiliation{Institute of Physics, Polish Academy of Sciences, Al. Lotnik\'ow 32/46, 02-668 Warsaw, Poland}

\date{\today}% It is always \today, today,

\begin{abstract}
The breakup of liquid threads
into smaller droplets is a 
fundamental problem in fluid dynamics. 
In this study, we estimate the 
characteristic wavelength of the breakup 
process by means of many-body dissipative
particle dynamics. This wavelength shows a power-law
dependence on the Ohnesorge number
in line with results from stability analysis.
We also discover that the number of satellite droplets
exhibits a power-law decay with 
exponent $0.72 \pm 0.04$ in the product 
of the Ohnesorge and thermal capillary numbers,
while the overall size of main droplets is larger
than that based on the characteristic wavelength thanks
to the asynchronous breakup of the thread.
Finally, we show that the formation of 
satellite droplets is the result of the advection of
pinching points towards
the main droplets in a remaining thinning neck, when
the velocity gradient of the fluid
exhibits two symmetric maxima.
\end{abstract}

\maketitle

\section{Introduction}
\begin{figure}[bt!]
\includegraphics[width=0.9\textwidth]{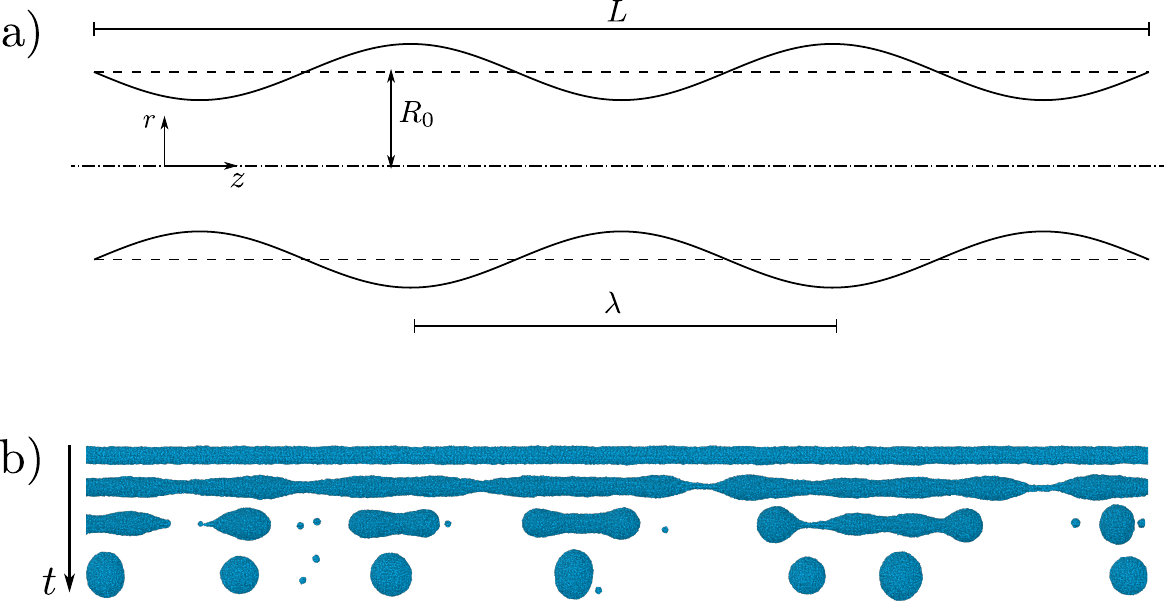}
\caption{\label{fig:1} a) Schematic of the initial configuration
of the system (dashed line) and the evolution of the
breakup of a liquid thread of an initial radius,
$R_{\rm 0}$. b) Characteristic snapshots from the
simulations during the time evolution of the breakup
from the liquid thread to individual droplets.
Here, only a part of the whole thread is shown 
for clarity, while $R_0=6$ 
\blu{(in reduced units throughout the paper)}.
}
\end{figure}

The breakup of liquid threads is not only a natural process 
observed on various occasions in everyday life
(\blu{e.g.,} breakup of a liquid thread when water 
falls from a tap),
\blu{but also relevant} for various applications \cite{Moseler2000}, such as
nanoprinting \cite{Basaran2013}, nanoscale manufacturing
and chemical processing \cite{Ye2003}, spraying \cite{wu2021},
and inkjet printing \cite{Hoath2016}.
The Rayleigh--Plateau instability \cite{Plateau1857,Rayleigh1878}, 
encountered during the pinching of liquid threads \cite{Eggers1997,Pita2015},
remains a fascinating phenomenon.
Many of its aspects require 
further investigations to reach a better understanding,
especially those related to its molecular origin,
which %can account for the 
\blu{significantly affects the singular} %\blu{molecular}  
behavior in the vicinity
of the \blu{breakup} %singular 
point. % around the breakup regime. 
According to stability analysis \cite{book:chandrasekhar,Plateau1857,Rayleigh1878}, 
the liquid thread becomes unstable and pinches off for any 
perturbation with wavelength larger 
than the unperturbed cylinder circumference, $2\pi R_0$ (Fig.~\ref{fig:1}) 
\cite{Plateau1849}. 
Moreover, linear
stability analysis predicts that the most unstable 
mode for an inviscid fluid occurs at reduced wavenumber $\chi=2\pi R_0/\lambda=0.697$ (the famous Rayleigh mode) \cite{Rayleigh1879},
which corresponds to a wavelength 
$\lambda\approx9.01R_{\rm 0}$ (Fig.~\ref{fig:1}a).
Plateau has estimated a wavelength of 
$4.38\times(2R_{\rm 0}$) \cite{Plateau1849}.
However, the exact value of this perturbation 
has not been properly quantified by molecular-scale \textit{in silico}
experiments \cite{tiwari2008,gopan2014}.
While the process is fundamentally driven by
surface tension, various parameters are expected to
affect this phenomenon, such as
inertial and viscous forces, and thermal fluctuations
\cite{ZhaoLockerbySprittles2020,Petit2012,Hennequin2006}.
Another aspect relates to the formation of satellite drops, generally
unfavorable for applications 
(\textit{e.g.}, inkjet printing). These
cannot be captured by a linear theory that would
predict a homogeneous breakup of the thread into equal
parts (c.f. snapshots of Fig.~\ref{fig:1}b). 
To properly describe the thread breakup
and understand the mechanisms of the formation
of satellite droplets,
a model that can take into account the thermal
fluctuations of the system needs to 
be employed.

In this study, we use a particle-based 
mesoscale model to uncover several important properties of  
the liquid-thread
breakup phenomenon. The model
properly accounts for fluid properties \cite{Ghoufi2011},
such as surface tension
and viscosity, and, also, includes thermal fluctuations
(see Section~\ref{model} for  details of the model and
simulation method). 
We determine the characteristic
wavelength of the formed structures during
the breakup of liquid threads of
different Ohnesorge (Oh) numbers and
juxtapose our predictions with
theoretical predictions
based on stability analysis \cite{book:chandrasekhar, Weber1931}.
We have also investigated
the formation of main and satellite droplets 
that occurs at longer times. It is found that 
the number of
satellite droplets follows a clear power-law dependence
on the Oh and \blu{thermal} capillary (Th) numbers that 
has not previously drawn attention, and that
the size of the main droplets is larger than what would 
be predicted based on the characteristic
wavelength at larger Oh numbers, due
to the asynchronous breakup of the thread. 
Finally, we show that the formation of 
satellite droplets is the result of the advection of
pinching points in the thin neck that remains 
between forming main droplets. 
The advection is due to the large
velocity gradients at the pinching points. 
Thus, we anticipate that our study sheds more light
into the underlying mechanisms of thread breakup
and its fundamental aspects.

\section{Model and Methodology}
\label{model}
The many-body dissipative particle
dynamics (MDPD) method \cite{,warren2003,zhao2017,zhao2021-review,zhao2021,zhao_sprittles_lockerby_2019,zhao2020} 
was used to carry out the simulations. 
\blu{This method can be used to describe flows 
at length and time scales relevant for describing
topological changes in flows, but at the same time %and 
naturally
handles singularities, such as the one developing 
at the pinching point during the breakup process, 
and provides
molecular-scale resolution.}
Fluids with different properties are studied, 
\blu{i.e.,} different Oh numbers, while the 
density, surface tension, and viscosity
of these fluids are the key properties and are
reported in Table~\ref{tab:values}.
The initial configuration of the system
consists of a cylindrical liquid thread with radius,
$R_{\rm 0}$,
and length, $L$, with periodic boundary conditions applied
at each end of the cylinder 
in the $z$ direction (Fig.~\ref{fig:1}). 
Characteristic snapshots obtained by the MDPD simulations
are also presented in Fig.~\ref{fig:1}b.
A range of relevant lengths, $L$, and radii, $R_{\rm 0}$, have
been considered to examine finite-size effects 
and gain better resolution on the properties studied,
for fluids \blu{characterized} by
different Oh numbers, defined as
$\text{Oh}=\mu/\sqrt{\rho\sigma R_{\rm 0}}$. 
Here, $\mu$ is the fluid's viscosity,
$\rho$ its density, and $\sigma$ its surface tension.
The thermal capillary number, which is also
relevant here, 
is defined as $\text{Th} = l_T/R_{\rm 0}$,
where $l_T$ is the thermal capillary length 
$l_T=\sqrt{k_BT/\sigma}$, %where 
$k_B$ is Boltzmann's constant
and $T$ the temperature of the system,
\blu{which are set to unity and define the 
energy scale of the model.}
The analysis of the results is based
on an ensemble of \textit{in silico} experiments for 
each set of parameters\blu{, in order} to obtain reliable
statistics.

The many-body dissipative particle dynamics (MDPD)
model consists in solving the Langevin equation 
of motion (Eq.~\ref{eq1} ) for each particle, $i$, 
that interacts with its neighbors, $j$, 
through a conservative force $\bm{F}^C$, \blu{i.e.},
\begin{eqnarray}
m\frac{d\bm{v}_i}{dt} = \sum_{j\neq i} \bm{F}_{ij}^C + \bm{F}_{ij}^R + \bm{F}_{ij}^D,
\label{eq1}
\end{eqnarray}
where $\bm{F}^R$ is a random force and 
$\bm{F}^D$ a dissipative force acting on each particle, $i$.
The main difference between \blu{the} MDPD and 
the standard DPD model in its most common formulation
is the expression for the conservative force, which in
the MDPD model reads
\begin{eqnarray}
\bm{F}^C_{ij} =  A\omega^C(r_{ij})\bm{e}_{ij} + 
B \left(\bar{\rho_i} + \bar{\rho_j} \right) \omega^d(r_{ij})\bm{e}_{ij},
\label{eq2}
\end{eqnarray}
where $A<0$ and $B>0$ are the attractive and repulsive
parameters, respectively, $r_{ij}$ is the 
distance between particles, $\bm{e}_{ij}$ is 
the unit vector in the direction
from particle $i$ to particle $j$, 
while $\omega^C(r_{ij})$ and $\omega^d(r_{ij})$
are linear weight functions, which are defined as follows:
\begin{eqnarray}
\omega^{C}(r_{ij}) = 
\begin{cases}
&1 - \frac{r_{ij}}{r_{c}}, \ \ r_{ij} \leq r_{c} \\
& 0,  \  \ r_{ij} > r_{c},
\end{cases} 
\label{eq3}
\end{eqnarray}
with $r_c$ being a cutoff distance for the interactions, 
usually set to unity. 
$\omega^d(r_{ij})$ has the same form,
however, its cutoff distance $r_d=0.75$, 
which is smaller than $r_c$.

\begin{table}[b]
\caption{\label{tab:values} Fluid properties obtained 
for different $A$ values in the MDPD model. 
The density, $\rho$, was obtained from the simulation, 
the surface tension, $\sigma^{\rm fit}$, 
and the viscosity, $\mu$, were calculated 
from fitting equations from Refs.~\citen{arienti2011}
and \citen{zhao2017}, respectively. Here,
$R_0=6$ was used to calculate Oh numbers. }
\begin{ruledtabular}
\begin{tabular}{ccccc}
 A  & $\rho$  & $\sigma^{fit}$  & $\mu$ & Oh  \\
 \hline 
-40   &   6.75    &   9.95    &   4.06    &   0.199 \\ 
-50   &   7.65    &   15.98   &   7.22    &   0.266   \\
-60   &   8.30    &   22.60   &   10.76   &   0.321   \\
-70   &   8.95    &   30.66   &   18.31   &   0.451   \\
-80   &   9.60    &   40.29   &   33.90   &   0.704   \\
-85   &   9.92    &   45.73   &   47.02   &   0.901   \\
-90   &   10.24   &   51.62   &   64.01   &   1.137   \\
\end{tabular}
\end{ruledtabular}
\end{table}

The repulsive term contains the many-body contributions
through its dependence on local neighborhood densities,
$\bar{\rho_i}$ and $\bar{\rho_j}$, 
which are calculated as follows:
\begin{eqnarray}
\bar{\rho_i} = 
\sum_{0<r_{ij}\le r_d}
\frac{105}{16\pi r_d^3} \left( 1 + 3 \frac{r_{ij}}{r_d}\right)
\left( 1 - \frac{r_{ij}}{r_d}\right)^3 .
\label{eq4}
\end{eqnarray}
The random and dissipative forces act as a thermostat in this model and are given by
\begin{eqnarray}
\bm{F}^D_{ij} = -\gamma \omega^D(r_{ij}) (\bm{e}_{ij} \cdot  \bm{v}_{ij})\bm{e}_{ij} ,
\label{eq5}
\end{eqnarray}
\begin{eqnarray}
\bm{F}^R_{ij} = \xi \omega^R(r_{ij}) \theta_{ij} \bm{e}_{ij} ,
\label{eq26}
\end{eqnarray}
where $\gamma$ is the dissipative strength, 
$\xi$ is the strength of the random force, 
$\bm{v}_{ij}$ is the relative velocity between particles, 
and $\theta_{ij}$ is a random 
variable from a Gaussian distribution with unit variance.
According to the fluctuation--dissipation theorem,
$\gamma$ and $\xi$ are related to each other by
\begin{eqnarray}
\gamma = \frac{\xi ^2}{2 k_B T}, 
\label{eq7}
\end{eqnarray}
where the temperature of the system is $T$ 
(set to $1$ in our units),
and the weight functions for the forces are
\begin{eqnarray}
\omega^D(r_{ij}) = \left[\omega^R(r_{ij})\right]^2 = 
\begin{cases}
\left( 1 - \frac{r_{ij}}{r_c}\right)^2,  & r_{ij} \leq r_{c} \\
 0,   & r_{ij} > r_{c},
\end{cases}
\label{eq8}
\end{eqnarray}
The time integration of the equation of motion 
is performed by using the modified velocity-Verlet algorithm
with a time step $\Delta t = 0.01$.

Different fluid properties were obtained in the simulations 
by means of the attractive parameter, $A$,
while keeping constant 
$B=25$, $r_c=1$, $r_d=0.75$ and $\gamma=18$. 
They are reported in Table~\ref{tab:values}. 
The density $\rho$ is the mean bulk bead density in 
simulation units, and should be distinguished from
the local values $\bar{\rho}_i$ in Eq.~\ref{eq4}.
The surface tension was obtained from the fit
proposed by Ref.~\citen{arienti2011} as
\begin{equation}
\sigma^{\rm fit} = -\frac{\pi}{240}\left( 0.42 A r_c^5 \rho^2 + 0.003 B r_d^5 \rho^3 \right)
\end{equation}
and the viscosity values were taken from the measurements in Ref.~\citen{zhao2017}.
The density values were directly measured 
during the initial stages of the simulation 
and they show slightly higher values 
than what is typically found in the literature by using the 
equation of state proposed 
by Ref.~\citen{warren2003} or \citen{jamali2015}, 
due to the difference in local density weight function used.

Simulation units can be related to physical units if 
desired, by matching density, temperature and further 
choosing appropriate values of the constants in the 
evolution equation to achieve the appropriate surface 
tension or other physical quantities for the substance 
under consideration --- as studied for example in Ref.~\citen{Ghoufi2011}. 
Moreover, since simulation units are related to physical 
units by a matching procedure, it is worth keeping in 
mind that changing one evolution parameter, such as $A$,
is equivalent to corresponding changes of dimensionless 
numbers in the fluid. For example, 
increasing attraction $|A|$ (as we have done in this study) 
increases density $\rho$ since the potential becomes 
shorter range, and also increases the surface tension
$\sigma^{\rm fit}$. Therefore, for example, the thermal 
capillary number Th, which quantifies the relative strength of thermal fluctuations, is reduced. Also, the shorter range of the inter-bead potential decreases the typical fluctuation range.

\begin{figure}[bt!]
\includegraphics[width=0.5\columnwidth]{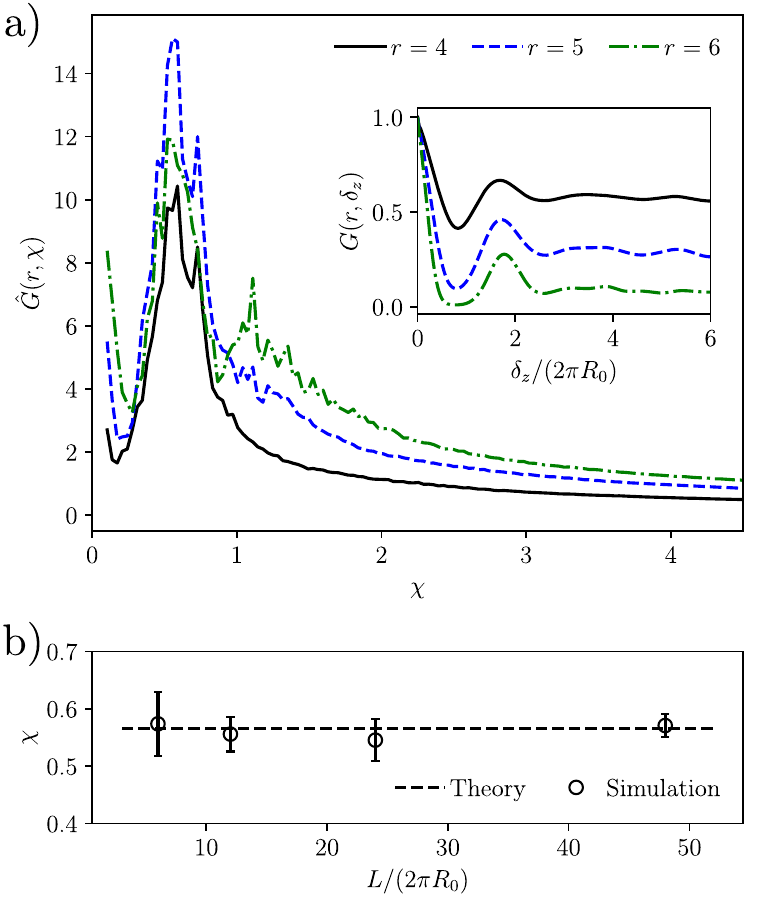}
\caption{\label{fig:2} a) Fourier
transform, $\hat{G}(r,q)$, of the correlation of
the density fluctuations, $G(r,\delta z)$ (inset),
at various radial distances, $r$, from
the thread axis, as indicated.
Example for $\text{Oh}=0.266$;
b) The characteristic
wavelength for threads of different lengths,
$L$.
The dashed line indicates the theoretical
prediction, $\chi = 0.565$, as obtained 
from Eq.~\ref{theory}.
}
\end{figure}

\section{Results and Discussion}
\label{results}
We have first
determined the characteristic wavelength of the
breakup process along the thread direction. 
To achieve this, we take advantage of the system symmetry
and use the correlation of the density fluctuations 
along the $z$ direction at radial distance, $r$, from
the cylindrical axis of the thread, which is expressed as
\begin{eqnarray}
G(r, \delta z) = \frac{\braket{\rho(r,\phi,z) \rho(r,\phi,z+\delta z)}_{z,\phi,T}}{\braket{\rho^2(r)}_{z,\phi, T}},
\end{eqnarray}
with $\rho(r,\phi,z)$ being the local density in a volume element
at radial distance, $r$, from the cylindrical
axis, while $\delta z \leq L/2$, due to the presence of
periodic boundary conditions in the $z$ direction \cite{Theodorakis_2009}. It is also
indicated that one takes the average over
an ensemble
of configurations at temperature, $T$,
over the angle, $\phi$,
oriented normal to the cylindrical axis, and over $z$.
The result of
the calculation provides the estimation of the characteristic
length scale, $\lambda$,
that develops during the thread breakup.
This length scale is more visible at larger distances from the axis, $r$,
but persists over a large range of $r$
(inset of Fig.~\ref{fig:2}a).
It can be determined from the peak
position $q_{\rm max}$ of the discrete Fourier transform
of $G(r, \delta z)$, as $\lambda=2\pi/q_{\rm max}$. Data for a particular case 
are shown in Fig.~\ref{fig:2}a.
A robust value of $q_{\rm max}$
is determined by a Gaussian fit to the points around
the peak.
Reliable statistics are 
obtained by \blu{realizing} an adequate ensemble of
simulations for each case,
while possible side effects due to the presence of
periodic boundary conditions have 
been investigated by considering
threads of different lengths, $L$.
Indeed, we have found that 
possible finite size effects quickly 
disappear as the length, $L$, becomes larger
than the circumference of the thread, $2\pi R_0$.
We henceforth consider
long threads to obtain the highest
possible accuracy on the characteristic wavelength through
the Fourier transform. Moreover, a larger number
of droplets formed in the case of longer
threads allows for better statistics 
on main and satellite droplet properties.

Figure~\ref{fig:2}b
\blu{summarizes} our results for the reduced 
characteristic (most unstable and smallest)
wavenumber $\chi = 2 \pi R_{\rm 0} / \lambda$.
We find that $\chi = 0.57\pm 0.05$
independently of the chosen 
length of the liquid thread, which is in agreement with 
previous predictions of stability analysis of
the Navier--Stokes equations 
(Eq.~\ref{theory}) \cite{book:chandrasekhar,Weber1931}. 
Moreover, breakup has not been observed in our simulations
when the length of the thread, $L$,
is smaller than the circumference of the thread,
in line with previous theoretical
arguments \cite{Plateau1849}.

\begin{figure}[bt!]
\includegraphics[width=0.6\columnwidth]{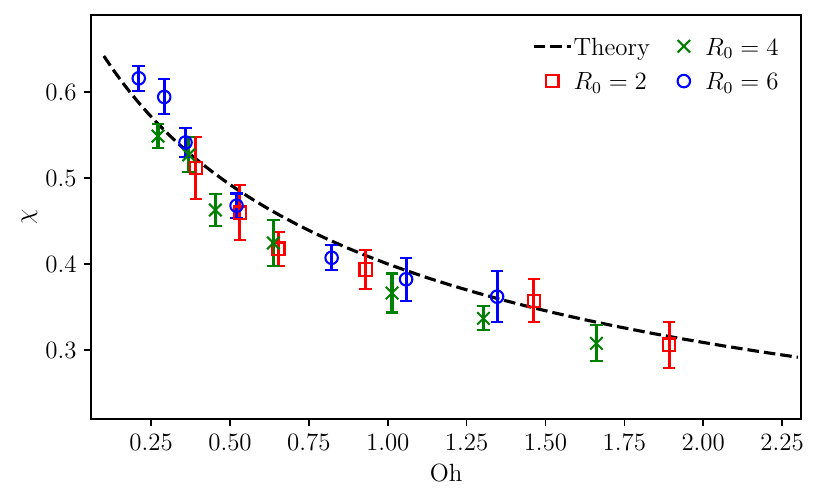}
\caption{\label{fig:3} Characteristic wavenumber,
$\chi$, \textit{versus} Oh number. The dashed line shows the theoretical
prediction (Eq.~\ref{theory}). 
}
\end{figure}

Fluids with different properties have been studied and the 
wavenumber, $\chi$, has been determined
for each case by using the same path as described 
above (Fig.~\ref{fig:2}). 
Then, Fig.~\ref{fig:3} presents the dependence
of $\chi$ on Oh (note that Oh depends not only on 
the fluid's properties, but also on $R_0$), 
which shows a very
good agreement with the predictions of
stability analysis \cite{book:chandrasekhar,Weber1931}.
According to the theory, the reduced wavelength
follows the relation
\begin{equation} \label{theory}
\chi = \sqrt{\frac{1}{2+\sqrt{18}\text{Oh}}}.
\end{equation}
Our results indicate that
Eq.~(\ref{theory}) is generally valid for threads
of different Oh in the range that can
be captured by the MDPD model,
which might suggest that $\chi$ has
a universal aspect and the stochastic
nature of breakup at different points
along the thread due to the thermal
fluctuations do not seem to affect the
wavenumber, $\chi$. In this context, note how decreasing 
$R_0$, as we have done while keeping fluid properties 
constant, increases the thermal capillary number Th and can 
be mapped to a system with higher $T$ and relatively more 
important thermal fluctuations.

\begin{figure}[bt!]
\includegraphics[width=\columnwidth]{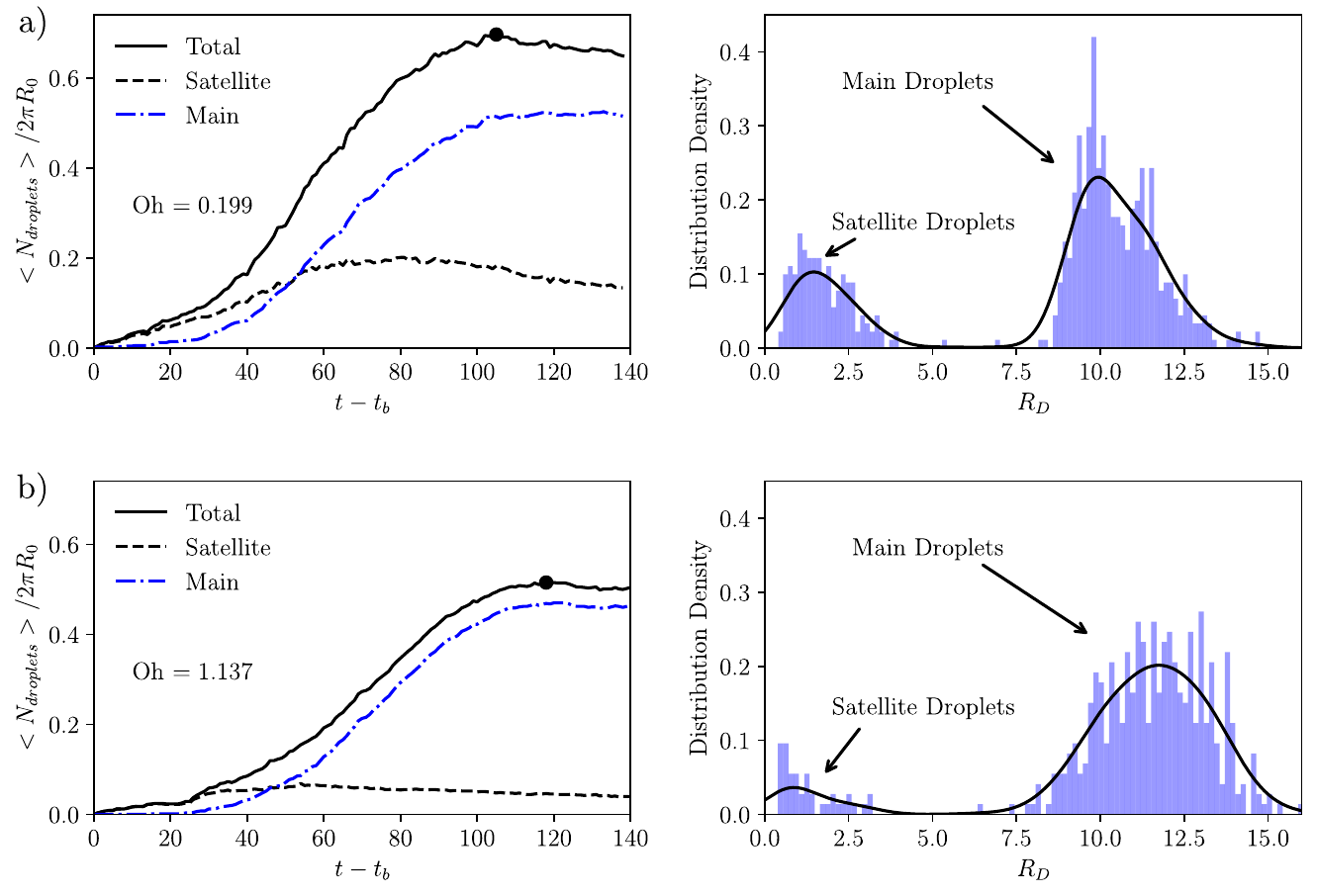}
\caption{\label{fig:4} The left panels illustrate the 
evolution of the linear number density of droplets \textit{versus}
time ($t_b$ is the time of the first breakup
event) for the main and satellite droplets, as well
the total sum of the two, as indicated, for 
a) $\textrm{Oh}=0.199$ and b) $\textrm{Oh}=1.137$.
The right panels show the distribution of the size
of droplets (calculated from the radius of gyration, $R_D$) 
at the time of maximum droplet count, 
indicated by the black dot in the left panels. 
The two distinct populations are the satellite 
(left peak at smaller $R_D$) and the main 
(right peak at larger $R_D$) droplets. 
Both cases had initial cylinder radius $R_0=6$.
}
\end{figure}

Since the individual breakup events take place at different
points along the liquid thread and times, the total number
of formed droplets varies during the simulation, and
there is a point in time when there is a maximum
number of droplets (Fig.~\ref{fig:4} shows examples
for higher and lower Oh-number cases).
To investigate their sizes and distribution, we have
used cluster analysis to identify the droplets,
where a distance $0.8$ between neighboring particles
has been used as the acceptance criterion to a cluster.
To consider a cluster as a fully formed
droplet (rather than remains of the thread), we require that its relative shape anisotropy \cite{Danilov2015}
\begin{equation}
    \kappa^2 = \frac{3}{2}\frac{\lambda_x^4+\lambda_y^4+\lambda_z^4}{\left(\lambda_x^2+\lambda_y^2+\lambda_z^2
    \right)^2} - \frac{1}{2}
\end{equation}
should satisfy the criterion that $\kappa^2 < 0.2$. 
Here, $\lambda_x$, $\lambda_y$ and $\lambda_z$ 
are the principal moments of the gyration tensor.
Values of relative shape anisotropy closer to $0$ indicate that a 
cluster has a stronger spherical symmetry 
while values closer to $1$ 
rather indicate that all points lie on a line. 

The time evolution of the linear density of the
number of droplets is shown in Fig.~\ref{fig:4}, 
along with the distribution of 
droplet sizes at the time of maximum droplet number.
We observe a majority population
of main droplets, characterized by a large
radius of gyration (obtained from $\lambda_{x,y,z}$),
and a smaller, very well separated, population
of satellite droplets with much smaller radii of gyration.
The reported average properties
that are related to the number of
droplets are calculated at this time 
of maximum linear density. 
The slow reduction in the number of droplets at later times
is due to subsequent coalescence
events, which after a long time should lead to
the formation of a single droplet encompassing
all particles.

\begin{figure}[bt!]
\includegraphics[width=0.6\columnwidth]{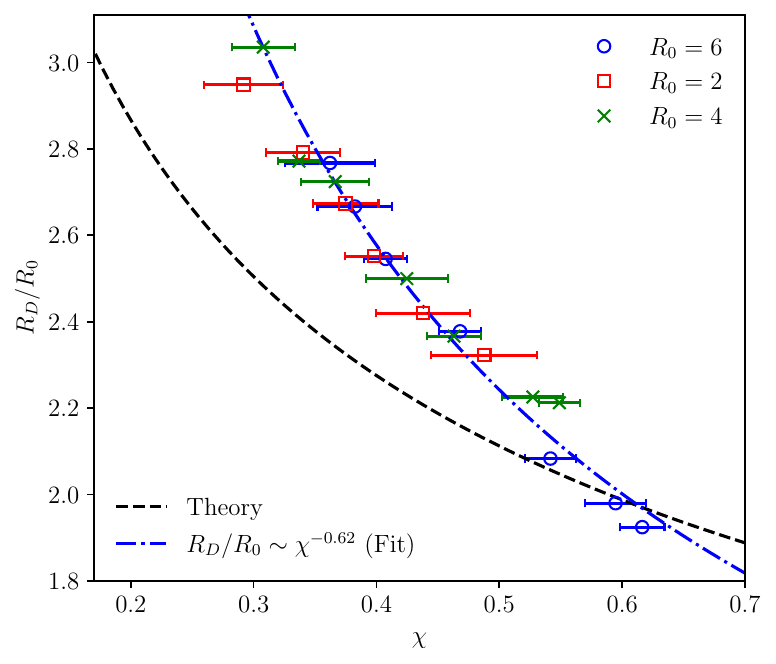}
\caption{\label{fig:5}  
Dependence of the average expected
radius, $R_{D}$, of the main droplets 
\textit{versus} the characteristic wavenumber, $\chi$.
The dashed line shows the expected value (Eq.~\ref{RD}) 
based on the 
volume balance with respect to $\chi$. 
\blu{The blue dashed line shows a power-law fit
of the simulation data, i.e., $R_D/R_0 \sim \chi^{-0.62}$.}
}
\end{figure}

\begin{figure}[bt!]
\includegraphics[width=\columnwidth]{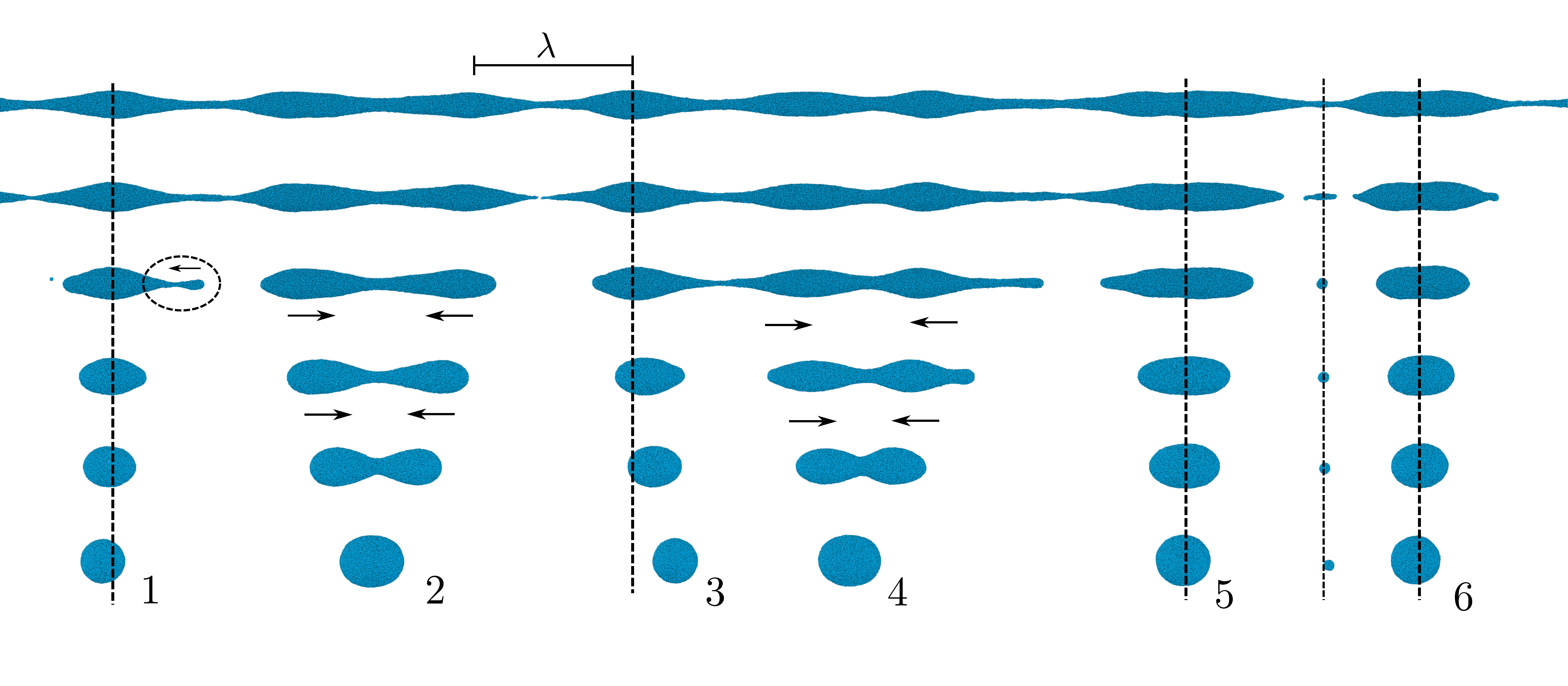}
\caption{\label{fig:6} Time evolution of a breakup simulation 
showing the mechanism of breakup suppression due to the
asymmetry in the pinching events,
which can also suppress the formation of satellite droplets,
and the mechanism by which
droplets acquire lateral momentum that will lead to
coalescence later on. Numbers indicate droplets discussed 
in the text. This simulation corresponds to the case where Oh $=1.137$ and 
$R_0=6$.
}
\end{figure}

\begin{figure}[bt!]
\centering
\includegraphics[width=.7\columnwidth]{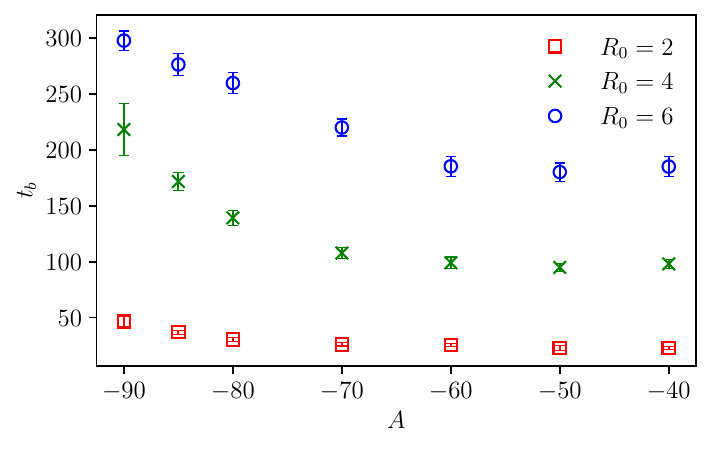}
\caption{\label{fig:7} Dependence of the breakup time $t_b$ on the attractive 
parameter $A$ for threads with different initial radius $R_0$. 
}
\end{figure}

\blu{Given that the characteristic wavelength $\chi$
naturally characterizes the breakup process, one
may 
estimate a theoretically expected droplet radius by assuming
%assume 
that the transformation of a part of the thread
with the initial cylindrical geometry of length
$\lambda$ and radius $R_0$ to a spherical droplet of radius
$R_D$ would take place without any loss of material. 
Hence, considering that the density of the liquid phase
does not change during the transformation from the cylinder to
the sphere, one can assume that $V_{\rm cyl}=V_{\rm d}$, 
where $V_{\rm cyl} = \pi R_{0}^{2} \lambda = 2 \pi^2 R_{0}^{3}/ \chi$ is the initial volume of the cylindrical 
part of the thread and 
$V_{\rm d}=4\pi R_{D}^{3}/3$ is the volume of the
formed droplet. Then, one %can obtain the 
obtains a theoretical expectation of the
relation 
between $R_D/R_0$ and $\chi$,  }
\begin{equation}\label{RD}
R_D/R_{\rm 0} = \sqrt[3]{3\pi / 2\chi}.
\end{equation}
For all fluids studied, 
we find that the average radius of the fully 
formed main droplets (Fig.~\ref{fig:5}), namely 
$R_D$ 
\blu{follows a downward trend qualitatively similar to the volume conserving expectation (\ref{RD}), but with a steeper ascent at small $\chi$. Thus}
%generally follows the trend of the theoretical expectation which 
%rather suggests that
$R_D$ \blu{does indeed} depend %s
on the wavenumber, $\chi$, but a %certain 
deviation from the \blu{simple} %theoretical 
expectation \blu{to larger drop sizes} is seen for
threads characterized by a larger Oh (smaller $\chi$).
\blu{We interpret this as being} %This is 
related to the asynchronous breakup of the thread 
combined with increased viscosity, 
which leads to the merging of isolated clusters of remnant 
neighboring droplets into larger ones than indicated by 
Eq.~\ref{RD}, instead of their breakup.
Freed of the pull of neighbors at the edges of a cluster, 
remnant precursors gain momentum towards the \blu{center} and merge. 
A clear example of the mechanism
is shown in Fig.~\ref{fig:6}.
Moreover, the same momentum gaining mechanism 
can inhibit the formation of satellite droplets.
Due to the longer times required for
breakup in the case of larger wavelengths
(Oh numbers, Fig.~\ref{fig:7})
the suppression of pinching points is facilitated,
and a larger deviation of the mean droplet
size from the theoretical expectation (Eq.~\ref{RD})
is seen.

In Figure~\ref{fig:6}, we show the time evolution of the breakup
of a section of a longer liquid thread in order to highlight the
reason that the average size of the droplets is larger than
the expected theoretical value (Eq.~\ref{RD}) 
that is directly linked with $\chi$.
Let's examine the formation of 
droplets $3$ and $4$ (Fig.~\ref{fig:6}).
In this case, the initial shape
of the liquid thread is
\blu{characterized} by three bulges of approximate
wavelength $\lambda=2\pi R_0/\chi$. 
These would appear to be precursors of three droplets. 
However, we can see that some droplets 
separate earlier and these may form smaller main droplets 
(1, 3, 5, 6), while other precursors coalesce with neighbors
before they can fully separate.
Thus, two domains in the same cluster that were 
initially heading to be droplets of size approximately
according to $\chi$ end up as one larger amalgamated droplet.
Such amalgamates form the larger droplets like 2 and 4. 
A more viscous fluid can be expected to give more
amalgamated cases, explaining the discrepancy 
seen in Fig.~\ref{fig:5} for higher Oh-number fluids. 
Early separation of droplets from the
rest of the thread also frees the precursors in-between 
from stretching forces. 
For example, separation of droplet $3$ from
the liquid thread that spans two characteristic
wavelengths to its right, frees the precursors of a force
pulling left, which leads from this moment to a visible 
acceleration inward within the cluster $4$ 
as shown by the arrows, and facilitates
the formation of just one droplet $4$.
The same mechanism can inhibit the formation of 
satellite droplets. For example, in the
region indicated by a dashed-line circle in Fig.~\ref{fig:6},
near droplet $1$, a satellite droplet precursor 
is reabsorbed into droplet $1$ after earlier separation of $2$.

The deviation between the actual mean size of the droplets and
the theoretical expectation based on the characteristic 
reduced wavenumber, $\chi$, is greater for lower values
of $\chi$, as can be seen in Fig.~\ref{fig:5}.
Moreover, Fig.~\ref{fig:3} clearly shows that smaller $\chi$
occurs for the higher Oh values. 
Hence, in Fig.~\ref{fig:7}, we plot the time required 
for the first breakup to occur depending 
on the attraction strength, 
and it is seen to be longer at strong attraction.
The higher Oh and viscosity here, as evidenced by
Table~\ref{tab:values}, lead to the
longer times required for the breakup to occur 
under these conditions.
Therefore the increased deviation of droplet size
can be attributed to the longer breakup 
timescales at low $\chi$, high Oh,
which allows more time for the amalgamation process
discussed above to occur and facilitates
the suppression of pinching points.

\begin{figure}[bt!]
\includegraphics[width=0.6\columnwidth]{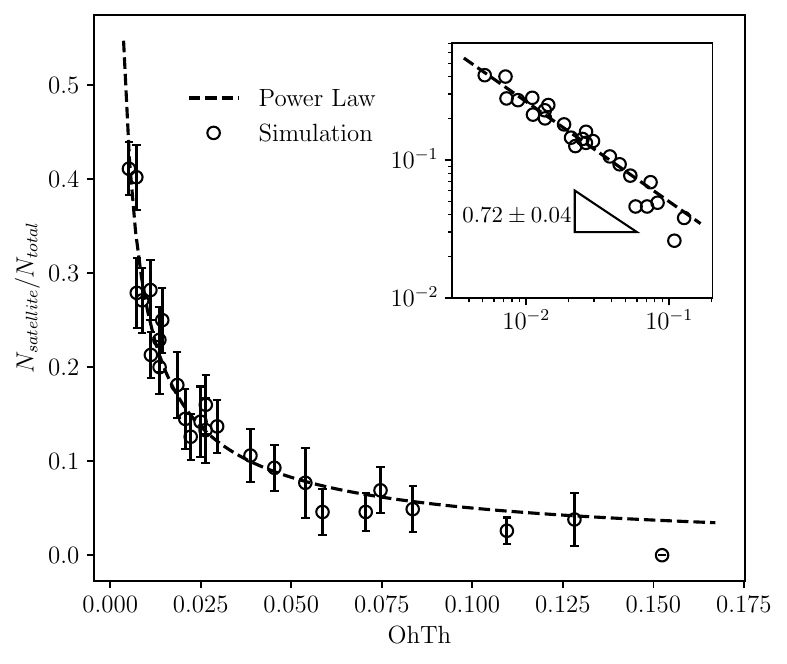}
\caption{\label{fig:8} Dependence of the proportion of satellite
droplets, $N_{satellite}$, compared to the total
number of droplets $N_{total}$, \textit{versus}
the product of the Ohnesorge and
thermal capillary numbers. 
}
\end{figure}

We have also counted the number of satellite
droplets, whose proportion exhibits
a power-law decrease
with increasing Oh and Th numbers
with exponent $-0.72 \pm 0.04$ as presented
in Fig.~\ref{fig:8}. 
While previous experimental results have suggested a 
linear dependence of $N_{satellite}/(N_{total}-N_{satellite})$
on the Th number \cite{Petit2012}, those have
not taken into account the additional dependence on
the Oh number. 
A significantly different relationship 
has been found here. 
Furthermore, after a certain limit, 
${\rm OhTh} \geq 0.15$, the thread breakup
does not yield satellite droplets any more.  
Unfortunately, we are not able to further probe
this limit with additional data for ${\rm OhTh} > 0.15$,
since in this case the MDPD model will yield a
solid phase, instead of the liquid one.

\begin{figure}[bt!]
\includegraphics[width=0.6\columnwidth]{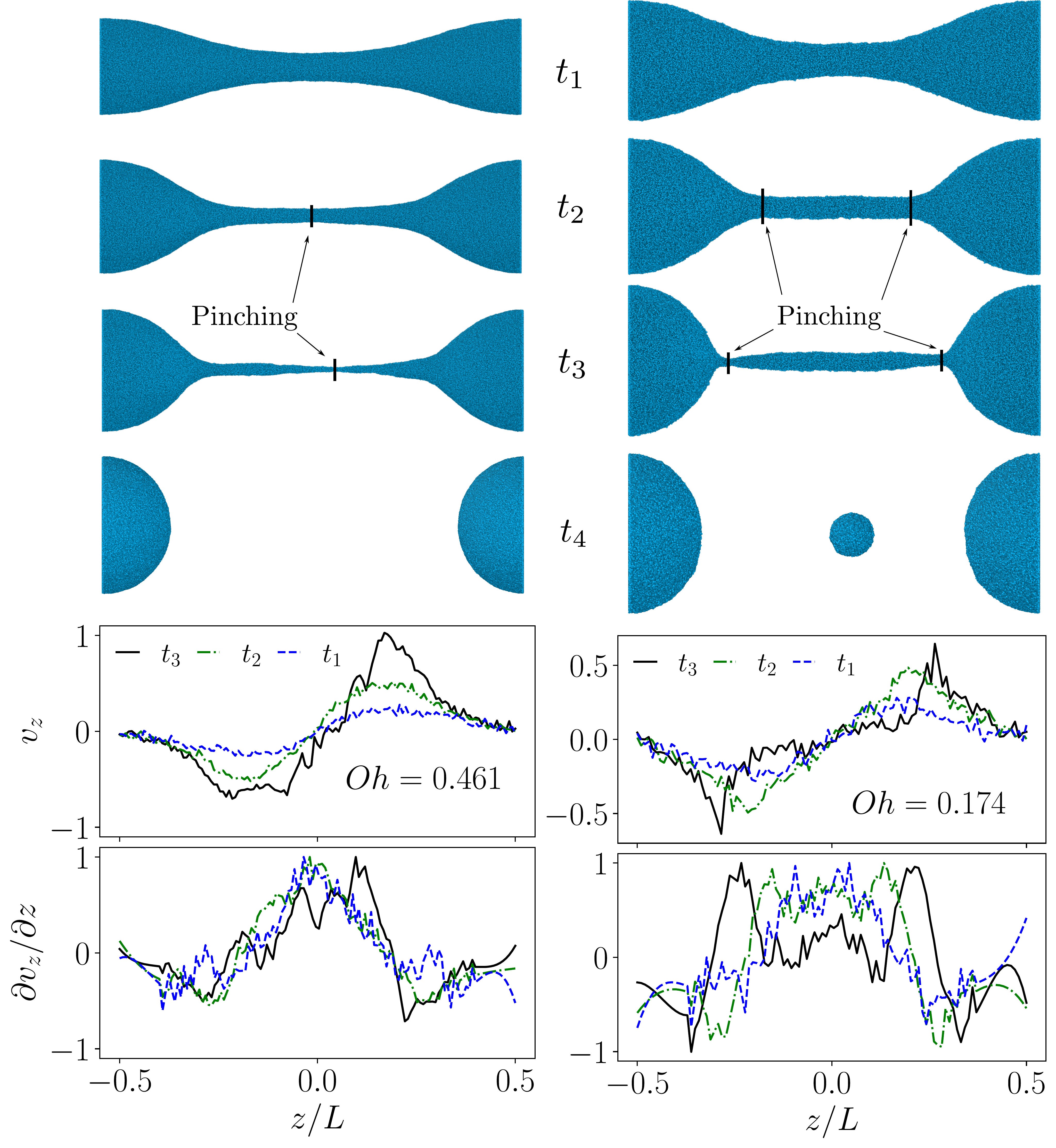}
\caption{\label{fig:9} Breakup of liquid threads
with different Oh numbers.
The formation of satellite droplets in
the case of lower $\rm Oh$ is seen and tracked in the right panel. 
We observe the advection of the 
pinching points towards the region that connects
the thinning neck to the main droplet. 
The advection coincides with maxima of the axial velocity gradient.
}
\end{figure}

It is important to try to understand the mechanism
of satellite droplet formation in more detail. 
%Theoretical studies
% have suggested that this might be due to the
% increased capillary pressure at the point
% where the thinning neck and the droplets
% connect \cite{martinez-calvo2020}. 
\blu{For this reason, we conducted simulations of a single breakup 
point under the perturbation of the most unstable wavenumber
by setting the cylinder length as $L=2 \pi R / \chi$ and
considering a case with a lower chance of satellite droplet 
formation (Oh = 0.461) and a case with higher chance (Oh = 0.174).
}
\blu{In turn}, we \blu{analyzed} the velocity field, $v_z$,
and its gradient, $\partial v_z / \partial z$, at
different times, $t$, during breakup \cite{martinez-calvo2020,Cohen1999,Cohen2001,Zhang1999}.
Figure~\ref{fig:9} shows the satellite-droplet formation sequence
and  the corresponding velocity field and its gradient. 
Tracking the evolution, we observe that satellite-droplet formation
occurs when two strong pinch points
advect towards the main droplets, as is 
seen on the right.
The figure also shows that relevant pinch points 
are associated with maxima of the velocity
\emph{and} of the velocity gradient, 
both of which shift in position.
For the satellite forming case (right), we see
that the velocity gradient profile is 
\blu{characterized} by two maxima at the pinching
points. In this case, 
the neck, which is the main part of the 
forming satellite droplet, 
cannot join either of the two main
droplets as the thinning process proceeds at
the pinch points. In contrast, 
in the case of larger Oh numbers, a single
pinching point will split the body of the neck
in two parts, which gradually join the main droplets.

\section{Conclusions}
\label{conclusions}
In this study, 
we have provided accurate values for the characteristic
wavenumber, $\chi$, for various fluids of
different Oh numbers by means of a 
particle-based mesoscale model. We find
that the dependence of $\chi$ on Oh is in good agreement
with the theoretical predictions based on 
stability analysis. 
Moreover, we have found that 
the number of satellite droplets follows
a very clear power-law decay with the product OhTh
with exponent $0.72 \pm 0.04$,
while the average main droplet size is
related with the characteristic wavelength 
\blu{characterizing} the breakup. However,
also, certain
deviations occur at larger Oh due to the
asynchronous breakup 
and increased timescales and viscosity that 
suppress the pinching points. We have also
found that satellite droplets form
when two approximately symmetric
peaks of the velocity distribution and its gradient,
which can correspond to a higher capillary 
pressure at the pinch points, 
move towards the ends of 
the thinning neck between the main  
droplets.
In such a scenario, the neck is not able to join either of
the two main droplets, thus forming a separate
satellite droplet. 
\blu{We anticipate that our results will motivate
further research in breakup phenomena, especially in the 
context of complex fluids, where MDPD, as a mesoscale
method, can offer significant advantages in describing 
macroscopic flows and at the same time providing
the detail required to capture relevant
molecular mechanisms in %due to 
the presence of 
various additives.}

\begin{acknowledgments}
This research has been supported by the National 
Science Centre, Poland, under
grant No.\ 2019/34/E/ST3/00232. 
We gratefully acknowledge Polish high-performance 
computing infrastructure PLGrid (HPC Centers: ACK Cyfronet AGH) 
for providing computer facilities and support 
within computational grant no. PLG/2022/015261.
\end{acknowledgments}

%\nocite{*}
%\bibliography{aipsamp}% Produces the bibliography via BibTeX.
\providecommand{\noopsort}[1]{}\providecommand{\singleletter}[1]{#1}%

\end{document}